\documentclass[twocolumn,showpacs,preprintnumbers,amsmath,amssymb]{revtex4}

\usepackage{graphicx}
\usepackage{dcolumn}
\usepackage{bm}

\begin{document}


\title{Experimental verification of a fully inseparable\\ 
tripartite continuous-variable state}

\author{Takao Aoki}%
 \email{takao@ap.t.u-tokyo.ac.jp}
\author{Nobuyuki Takei}
\author{Hidehiro Yonezawa}
\author{Kentaro Wakui}
\author{Takuji Hiraoka}
\author{Akira Furusawa}
\affiliation{%
Department of Applied Physics, School of Engineering, 
The University of Tokyo, 7-3-1 Hongo, Bunkyo-ku, Tokyo 113-8656, Japan
}%

\author{Peter van Loock}
\affiliation{
Quantum Information Theory Group, Zentrum f\"{u}r 
Moderne Optik, Universit\"{a}t Erlangen-N\"{u}rnberg, 91058 Erlangen, Germany
}%

\date{\today}

\begin{abstract}
A continuous-variable tripartite entangled state is experimentally generated 
by combining three independent squeezed vacuum states and the variances 
of its relative positions and total momentum are measured. We show that 
the measured values violate the separability criteria based on the sum of 
these quantities and prove the full inseparability of the generated state.
\end{abstract}


\maketitle
The remarkable proposal of quantum teleportation \cite{Bennett93} 
demonstrates that the quantum correlations of a shared entangled state 
enable two parties to reliably exchange quantum information.
So far, several experiments on quantum communication with 
discrete-variable states have been carried out.
In the domain of continuous variables (CVs), the unconditional 
quantum teleportation of arbitrary coherent states %
\cite{Furusawa98a,Bowen03a,Zhang03a} and quantum dense coding \cite{Li02a} 
have been demonstrated.
These successful experiments show the advantage of 
CV bipartite entanglement for the implementation of quantum protocols; 
that is, the simplicity of its generation and manipulation 
and the applicability of efficient homodyne techniques
to its detection.

CV entanglement may also be applicable to quantum protocols
involving more than two parties.
For example, tripartite entanglement 
(the entanglement shared by three parties)
enables one to construct a quantum 
teleportation network \cite{Loock00a}, to build an optimal one 
to two telecloner \cite{Loock01a}, or to perform controlled dense 
coding \cite{Zhang02a}.
CV tripartite entanglement can be generated in a similar way 
as in the case of CV bipartite entanglement.
It only requires combining 
three modes using linear optics, where at least one of these modes 
is in a squeezed state \cite{Loock00a}. In fact, as pointed out in 
Ref.~\cite{Giedke01a}, CV tripartite entanglement has already been 
generated in the CV quantum teleportation experiment
of Ref.~\cite{Furusawa98a}, 
although no further investigation was made there. 
On the other hand, the separability properties of tripartite states are 
more complicated than in the bipartite case; three-mode Gaussian states 
are classified into five different classes \cite{Giedke01a}. 
In order to exploit the tripartite entanglement 
for three-party quantum protocols such as that from Ref.~\cite{Loock00a}, 
the state involved has to be 
fully inseparable (class 1 in Ref.~\cite{Giedke01a}). 
Although the output state that emerges
from the beam splitters with one or more 
squeezed input states is in principle fully inseparable
for any nonzero squeezing \cite{Loock00a}, 
inevitable losses in the real experiment may destroy
the genuine tripartite entanglement and convert the state
into a partially or fully separable one.
This would make a true tripartite quantum protocol fail.
In other words, the success of a true 
tripartite quantum protocol ({\it e.g.},
a coherent-state quantum teleportation network with fidelities
better than one half)
is a sufficient criterion for the full
inseparability of the state involved \cite{Loock00a}. 
It should be noted here that the success of 
a tripartite quantum protocol between two parties with the help of 
the third party ({\it e.g.}, via a momentum detection of the mode three)
does not guarantee that the third party is inseparable from the rest.
In the example of the protocol of Ref.~\cite{Loock00a}, the full 
inseparability can be proven only when the protocol succeeds between 
at least two different pairs or, more generally, when the positions 
and momenta of all three parties are part of the protocol.

Though the full inseparability is unambiguously verified this way, 
an alternative verification scheme that does not rely on a full
quantum protocol is desirable. 
In the bipartite case, the inseparability may also be verified simply 
by measuring the variances of relative position 
and total momentum \cite{Duan00a, Simon00a}. 
Recently, a similar scheme to verify the full inseparability 
of CV tripartite entangled states was proposed \cite{Loock02a},
based on the variances of appropriate linear combinations
in position and momentum.
In this letter, we generate a tripartite 
entangled state by combining three independent squeezed vacuum states 
and demonstrate its full inseparability by applying the scheme
of Ref.~\cite{Loock02a}. 

Let us introduce the position and momentum 
quadrature-phase amplitude operators $\hat{x}$ and $\hat{p}$ 
corresponding to the real and imaginary part of an electromagnetic field 
mode's annihilation operator, respectively:
$\hat{a} = \hat{x} + i \hat{p}$ (units-free with $\hbar = \frac{1}{2}$, 
$[\hat{x} , \hat{p}] = \frac{i}{2}$). 
The simplest way to generate a tripartite entangled state is to send 
a single-mode squeezed vacuum state $|x = 0 \rangle $ 
(idealized by an eigenstate corresponding to infinite squeezing)
into a series of two beam splitters \cite{Loock00a}. 
In this case, the inputs of the two unused ports are vacuum states.
This is practically easy to implement, but when applied to a 
quantum protocol, the performance would be of only limited quality
due to the two vacuum input states.
For example, in the teleportation network, the maximum fidelity 
between any pair is then $1/\sqrt{2}$ in the limit of infinite squeezing 
(excluding additional local squeezers). 
In order to approach unit fidelity (perfect teleportation), 
one needs to send squeezed states into all input ports. 
An example is 
the CV counterpart of the Greenberger-Horne-Zeilinger (GHZ) 
state \cite{Greenberger90a}, 
$\int \!\! dx \, |x \rangle_1 |x \rangle_2 |x \rangle_3 $.
This CV GHZ state can be generated by sending 
a momentum-squeezed vacuum state $|p=0 \rangle_1 $ 
and two position-squeezed vacuum states $|x=0 \rangle_2 $ and 
$|x=0 \rangle_3 $ into a ``tritter'' \cite{Braunstein98a}, 
which consists of two beam splitters with 
transmittance/reflectivity of 1/2 and 1/1.
In order to show this, we define a beam splitter operator 
$\hat{B}_{ij}(\theta)$ which transforms two input modes $\hat{a}_{i,j}$ as
\begin{eqnarray}
\hat{B}^{\dag}_{ij}(\theta)
\left(
\begin{array}{c}
\hat{a}_i \\
\hat{a}_j 
\end{array}
\right)
\hat{B}_{ij}(\theta)
=
\left(
\begin{array}{c}
\hat{a}_i \cos \theta + \hat{a}_j \sin \theta \\
\hat{a}_i \sin \theta - \hat{a}_j \cos \theta 
\end{array}
\right).
\end{eqnarray}
The transmittance $T$ and the reflectivity $R$ of the beam splitter are
expressed by $T = \cos^2 \theta$ and $R = \sin^2 \theta$, respectively.
Applying first 
$\hat{B}_{12}(\cos^{-1} 1/\sqrt{3})$ and then 
$\hat{B}_{23}(\pi/4)$ to the input state 
$|p=0 \rangle_1 |x=0 \rangle_2 |x=0 \rangle_3 $ yields 
$\int \!\! dx \, |x \rangle_1 |x \rangle_2 |x \rangle_3 $.
This CV GHZ state is a simultaneous eigenstate of zero
total momentum ($p_1+p_2+p_3 = 0$) and zero relative positions 
($x_i-x_j = 0$) and exhibits maximal entanglement.

In the real experiment, only finite squeezing is available.
Thus the output state is no longer the ideal CV GHZ state and 
it can never be maximally entangled. 
Accordingly, total momentum and relative positions have finite variances: 
$\langle [\Delta (\hat{p}_1 + \hat{p}_2 + \hat{p}_3)]^2 \rangle > 0$ and 
$\langle [\Delta (\hat{x}_i - \hat{x}_j)]^2 \rangle > 0$.
This becomes clear when we express the operators for the three output modes 
in the Heisenberg picture \cite{Loock00a}: 
\begin{eqnarray}
\hat{x}_1 &=& \frac{1}{\sqrt{3}} e^{+r_1} \hat{x}_1^{(0)} 
           +  \sqrt{\frac{2}{3}} e^{-r_2} \hat{x}_2^{(0)}, 
\nonumber \\
\hat{p}_1 &=& \frac{1}{\sqrt{3}} e^{-r_1} \hat{p}_1^{(0)} 
           +  \sqrt{\frac{2}{3}} e^{+r_2} \hat{p}_2^{(0)}, 
\nonumber \\
\hat{x}_2 &=& \frac{1}{\sqrt{3}} e^{+r_1} \hat{x}_1^{(0)} 
           -  \frac{1}{\sqrt{6}} e^{-r_2} \hat{x}_2^{(0)} 
           +  \frac{1}{\sqrt{2}} e^{-r_3} \hat{x}_3^{(0)}, 
\nonumber \\
\hat{p}_2 &=& \frac{1}{\sqrt{3}} e^{-r_1} \hat{p}_1^{(0)} 
           -  \frac{1}{\sqrt{6}} e^{+r_2} \hat{p}_2^{(0)} 
           +  \frac{1}{\sqrt{2}} e^{+r_3} \hat{p}_3^{(0)}, 
\nonumber \\
\hat{x}_3 &=& \frac{1}{\sqrt{3}} e^{+r_1} \hat{x}_1^{(0)} 
           -  \frac{1}{\sqrt{6}} e^{-r_2} \hat{x}_2^{(0)} 
           -  \frac{1}{\sqrt{2}} e^{-r_3} \hat{x}_3^{(0)}, 
\nonumber \\
\hat{p}_3 &=& \frac{1}{\sqrt{3}} e^{-r_1} \hat{p}_1^{(0)} 
           -  \frac{1}{\sqrt{6}} e^{+r_2} \hat{p}_2^{(0)} 
           -  \frac{1}{\sqrt{2}} e^{+r_3} \hat{p}_3^{(0)}. 
\label{GHZeq}
\end{eqnarray}
Here a superscript $(0)$ denotes initial vacuum modes, and $r_1$, 
$r_2$, and $r_3$ are the squeezing parameters. 
In addition to the finite squeezing, the inevitable losses 
in the experiment further degrade the entanglement. 
It is important to stabilize the relative phase 
of the three input modes
in order to properly adjust the squeezing directions.
The phase fluctuations in this stabilization lead to an extra
degradation of the entanglement.
As a result, the output state does not necessarily
exhibit genuine tripartite 
entanglement: it may be fully or partially separable. 
Therefore, we need to experimentally verify the full inseparability of 
the state.

A feasible scheme for this purpose is to check the following 
set of inequalities \cite{Loock02a}:
\begin{eqnarray}
{\rm I}. \quad
\langle [\Delta (\hat{x}_1 - \hat{x}_2)]^2 \rangle
+
\langle [\Delta (\hat{p}_1 + \hat{p}_2 + g_3\hat{p}_3)]^2 \rangle
\geq 1,
\nonumber \\
{\rm II}. \quad
\langle [\Delta (\hat{x}_2 - \hat{x}_3)]^2 \rangle
+
\langle [\Delta (g_1\hat{p}_1 + \hat{p}_2 + \hat{p}_3)]^2 \rangle
\geq 1,
\nonumber \\
{\rm III}. \quad
\langle [\Delta (\hat{x}_3 - \hat{x}_1)]^2 \rangle
+
\langle [\Delta (\hat{p}_1 + g_2\hat{p}_2 + \hat{p}_3)]^2 \rangle
\geq 1.
\label{ineq}
\end{eqnarray}
Here, the $g_i$ are arbitrary real parameters. Note that the variances of the
vacuum state are
$\langle (\Delta \hat{x}_i^{(0)})^2 \rangle =
\langle (\Delta \hat{p}_i^{(0)})^2 \rangle =
\frac{1}{4}$.
The violation of inequality ${\rm I}.$ is a sufficient condition for
the inseparability of modes 1 and 2 and is a criterion for the success of a
quantum protocol between parties 1 and 2.
Note that inequality ${\rm I}.$ alone
does not impose any restriction on the separability of mode 3 from the others.
In other words, the success of a quantum protocol between parties
1 and 2 with the help of party 3 (by conveying classical information
about a measurement of $\hat{p}_3$ \cite{Loock00a})
does not prove the inseparability of the third party from the rest.
Thus, we need to check the violation of at least two of the three
inequalities (\ref{ineq}) to verify the full inseparability 
of the tripartite entangled state \cite{Loock02a}.

From Eq. (\ref{GHZeq}) we find that the optimum gain $g_i^{\rm opt}$ 
to minimize the l.h.s. of Ineq. (\ref{ineq}) depends on 
the squeezing parameters, namely
\begin{eqnarray}
g_i^{\rm opt} =
\frac{e^{+2r_2}-e^{-2r_1}}{e^{+2r_2}+\frac{1}{2}e^{-2r_1}},
\label{opt} 
\end{eqnarray}
where $r_2=r_3$ (which makes the three-mode state totally symmetric
and hence $g_i^{\rm opt}$ independent of $i$).
In the case of infinite squeezing (CV GHZ state),
the optimum gain $g_i^{\rm opt}$ is one, while it is less than one 
for finite squeezing. Although the smallest values of the l.h.s. of
Ineq. (\ref{ineq}) are observed when we experimentally adjust 
$g_i^{\rm opt}$, we employ $g_i = 1$ for all $i$.
This makes the experimental verification simpler.
Moreover, the measured variances then directly correspond to those
of the eigenvalues of the ideal CV GHZ state
(relative positions and total momentum).

\begin{figure}
\resizebox{8cm}{!}{\includegraphics{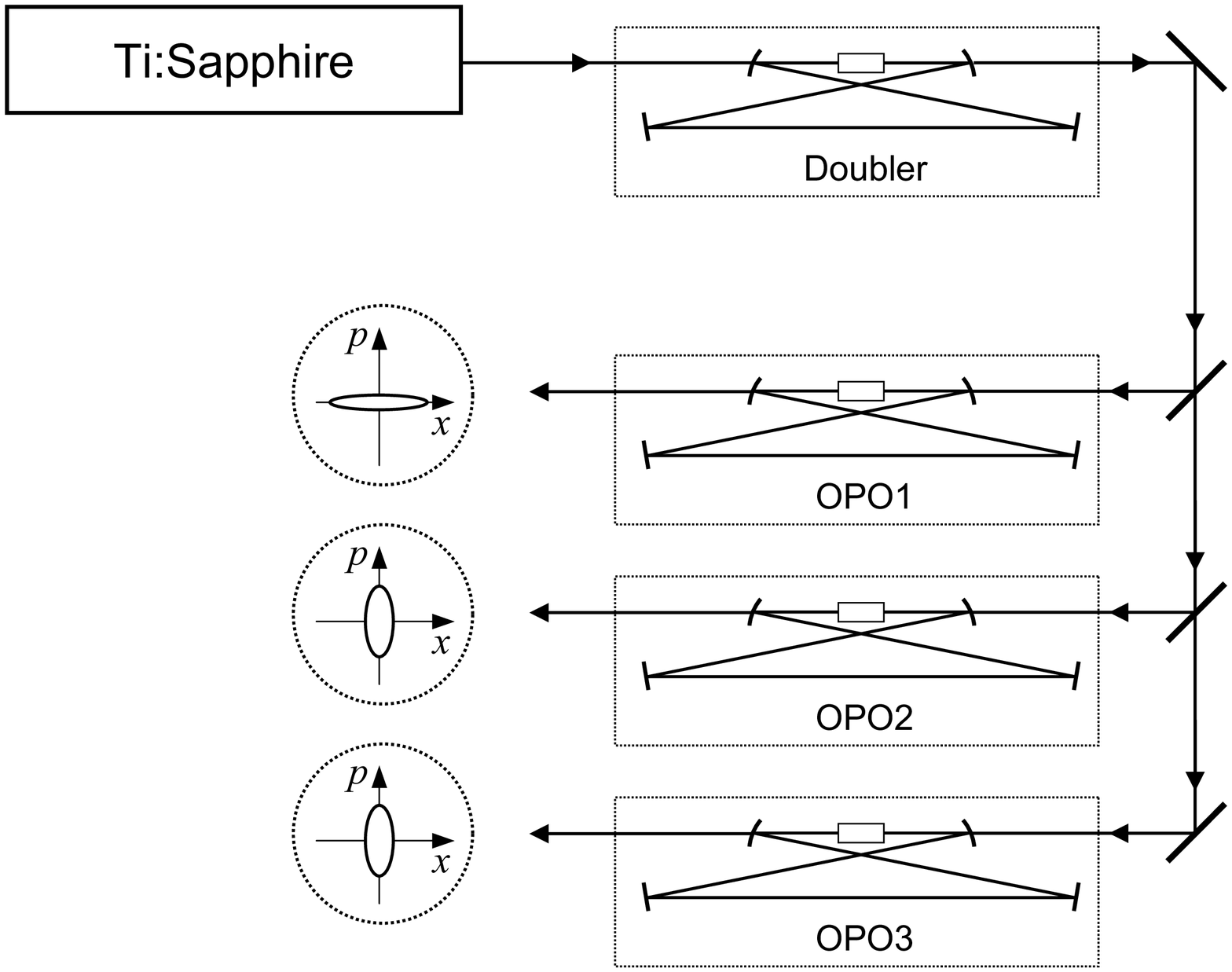}}
\caption{\label{f0} Schematic of the generation of three independent 
squeezed vacuum states.
}
\end{figure}

\begin{figure}
\resizebox{8cm}{!}{\includegraphics{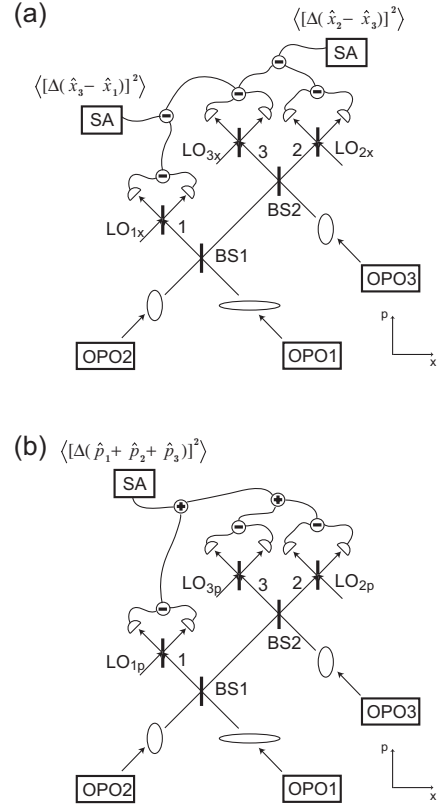}}
\caption{\label{f1} Schematic of the measurements of the variances 
(a) $\langle [\Delta (\hat{x}_3 - \hat{x}_1)]^2 \rangle $ and 
$\langle [\Delta (\hat{x}_2 - \hat{x}_3)]^2 \rangle $ and (b) 
$\langle [\Delta (\hat{p}_1 + \hat{p}_2 + \hat{p}_3)]^2 \rangle $.
BS1 and BS2 are beam splitters with $T/R$
ratios of 1/2 and 1/1, respectively.
The ellipses illustrate the squeezed quadrature of each beam.
LO$_{ix,p}$ denote local oscillator beams for homodyne detector $i$
with their phases locked at the $x$ and $p$ quadratures, respectively. }
\end{figure}

Figure \ref{f0} shows the schematic of the experimental setup to generate 
three independent squeezed vacuum states.
We use a subthreshold degenerate optical parametric oscillator (OPO) 
with a potassium niobate crystal (length 10mm). 
Each OPO cavity is a bow-tie type ring cavity which consists of two 
spherical mirrors (radius of curvature 50mm) and two flat mirrors.
The round trip length is 500mm and the waist size in the crystal is 
20$\mu$m.
An output of a Ti:Sapphire laser at 860nm is frequency-doubled in an external 
cavity with the same configuration as
for the OPOs and divided into three beams
to pump three OPOs. The pump powers are 56, 71, and 78 mW for 
OPO 1, 2, and 3, respectively.
The squeezed vacuum outputs from these OPOs are combined at two beam splitters
to generate the approximate CV GHZ state (see Fig.~\ref{f1}).
The visibilities of this combination are 0.968 for the input modes
1 and 2, and 0.948 for 2 and 3.
The output modes from the beam splitters are fed into the homodyne detectors
1, 2, and 3 with local oscillator (LO) powers of 1.3, 1.7, and 1.5 mW, 
and visibilities between the input modes to the homodyne detectors and LOs
of 0.979, 0.971, and 0.989, respectively.

\begin{figure}
\resizebox{8cm}{!}{\includegraphics{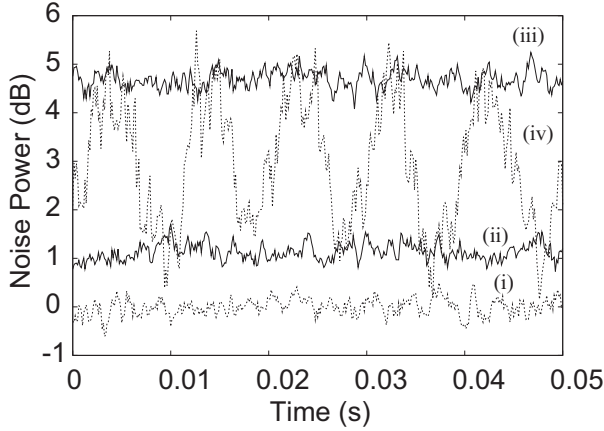}}
\caption{\label{f2} Noise measurement results on output mode 1 alone. (i) 
represents the corresponding vacuum noise 
$\langle (\Delta \hat{x}_1^{(0)})^2 \rangle = \frac{1}{4} $; 
(ii) the noise of the $x$ quadrature $\langle (\Delta \hat{x}_1)^2 \rangle $;
(iii) the noise of the $p$ quadrature $\langle (\Delta \hat{p}_1)^2 \rangle $;
(iv) the noise of the scanned phase. The measurement frequency is centered at
900kHz, resolution bandwidth is 30kHz, video bandwidth is 300Hz. Except for 
(iv) traces are averaged ten times.}
\end{figure}

We first measure the noise power of each output mode. 
Figure \ref{f2} shows the measurement results on output mode 1. 
The minimum noise level of 1.14$\pm$0.25 dB 
compared to the corresponding vacuum noise level is observed 
for the $x$ quadrature, while the maximum noise level of 4.69$\pm$0.26 dB 
is observed for the $p$ quadrature. 
Similarly, the minimum noise levels of 0.75$\pm$0.27 and
1.21$\pm$0.29 dB for the $x$ quadrature and the maximum noise levels 
of 4.12$\pm$0.27 and 4.69$\pm$0.21 dB for $p$
are observed for output modes 2 and 3, respectively.
Note that the observed noise levels are always above
the corresponding vacuum noise level.

Next we measure the variances of the relative positions
and the total momentum from Ineq. (\ref{ineq}).
Figure \ref{f1}(a) shows the schematic of the measurement of the variances 
$\langle [\Delta (\hat{x}_3 - \hat{x}_1)]^2 \rangle $ and 
$\langle [\Delta (\hat{x}_2 - \hat{x}_3)]^2 \rangle $.
The outputs of the homodyne detection are electronically subtracted, 
and the noise power is measured by spectrum analyzers.
The variance $\langle [\Delta (\hat{x}_1 - \hat{x}_2)]^2 \rangle $
is measured in a similar manner.
In the case of the variance 
$\langle [\Delta (\hat{p}_1 + \hat{p}_2 + \hat{p}_3)]^2 \rangle $, the 
noise power of the electronical sum of the homodyne detection outputs 
is measured as shown in Fig.~\ref{f1}(b). 

\begin{figure}
\resizebox{8cm}{!}{\includegraphics{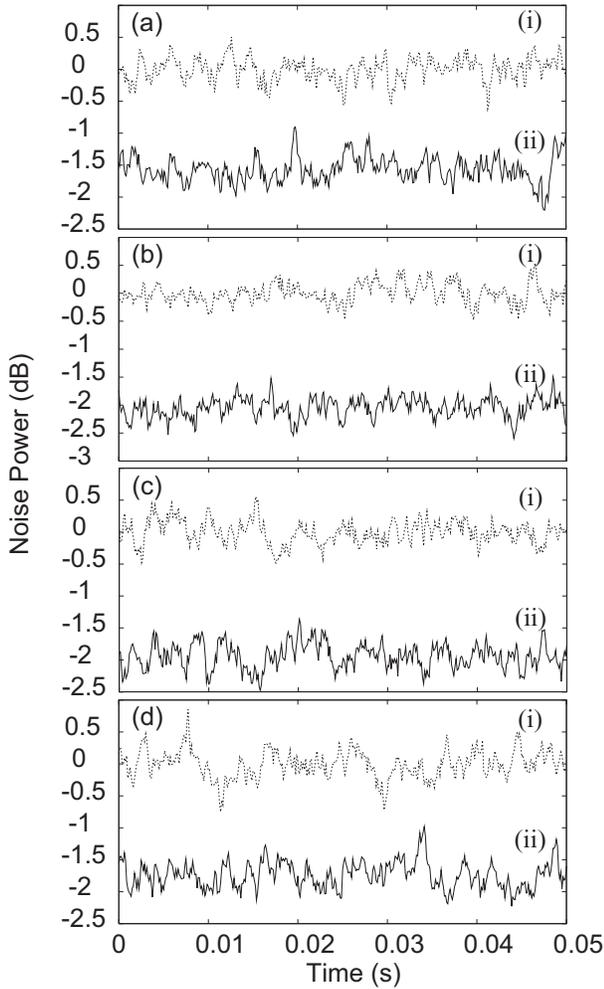}}
\caption{\label{f3} Noise measurement results corresponding to the variances 
of the l.h.s. of Ineq. (\ref{ineq}).
(a) 
(i) is $\langle [\Delta (\hat{x}_1^{(0)} - \hat{x}_2^{(0)})]^2 \rangle 
= \frac{1}{2}$ and 
(ii) is $\langle [\Delta (\hat{x}_1 - \hat{x}_2)]^2 \rangle $;
(b) 
(i) $\langle [\Delta (\hat{x}_2^{(0)} - \hat{x}_3^{(0)})]^2 \rangle 
= \frac{1}{2}$ and 
(ii) $\langle [\Delta (\hat{x}_2 - \hat{x}_3)]^2 \rangle $;
(c) 
(i) $\langle [\Delta (\hat{x}_3^{(0)} - \hat{x}_1^{(0)})]^2 \rangle 
= \frac{1}{2}$ and 
(ii) $\langle [\Delta (\hat{x}_3 - \hat{x}_1)]^2 \rangle $;
(d)
(i) $\langle [\Delta (\hat{p}_1^{(0)} + \hat{p}_2^{(0)} + \hat{p}_3^{(0)})]^2 
\rangle = \frac{3}{4}$ and 
(ii) $\langle [\Delta (\hat{p}_1 + \hat{p}_2 + \hat{p}_3)]^2 \rangle $.
The measurement conditions are the same as for Fig.~\ref{f2} with ten times averages.
}
\end{figure}

Figure \ref{f3} shows a series of measurement results of
(a) $\langle [\Delta (\hat{x}_1 - \hat{x}_2)]^2 \rangle $, 
(b) $\langle [\Delta (\hat{x}_2 - \hat{x}_3)]^2 \rangle $, 
(c) $\langle [\Delta (\hat{x}_3 - \hat{x}_1)]^2 \rangle $, 
and (d) $\langle [\Delta (\hat{p}_1 + \hat{p}_2 + \hat{p}_3)]^2 \rangle $, 
which have the average noise power of $-1.95$, $-2.04$, $-1.78$, 
and $-1.75$ dB, respectively, 
compared to the corresponding vacuum noise level. 
These results clearly show the nonclassical correlations among the three modes.
After repeating the measurement series ten times, we obtain the
following measured
values for the l.h.s. of Ineq. (\ref{ineq}),
\begin{eqnarray}
{\rm I}. \quad
\langle [\Delta (\hat{x}_1 - \hat{x}_2)]^2 \rangle
+
\langle [\Delta (\hat{p}_1 + \hat{p}_2 + \hat{p}_3)]^2 \rangle
\nonumber \\
= 0.851 \pm 0.062 < 1,
\nonumber \\
{\rm II}. \quad
\langle [\Delta (\hat{x}_2 - \hat{x}_3)]^2 \rangle
+
\langle [\Delta (\hat{p}_1 + \hat{p}_2 + \hat{p}_3)]^2 \rangle
\nonumber \\
= 0.840 \pm 0.065 < 1,
\nonumber \\
{\rm III}. \quad
\langle [\Delta (\hat{x}_3 - \hat{x}_1)]^2 \rangle
+
\langle [\Delta (\hat{p}_1 + \hat{p}_2 + \hat{p}_3)]^2 \rangle
\nonumber \\
= 0.867 \pm 0.062 < 1.
\label{ineqresult}
\end{eqnarray}
Since violations of all the inequalities are demonstrated, we have proven the
full inseparability of the generated tripartite entangled state. 

In summary, we have generated a tripartite CV entangled state
and verified its full inseparability according to the criteria
based on the variances of the relative positions and the total momentum.
The violations of all the inequalities
verify the presence of genuine tripartite entanglement.
Moreover, they ensure that a suitable true tripartite quantum 
communication protocol using the generated state would succeed.
For example, a fidelity greater than one half would be achievable 
between any pair of parties in a tripartite quantum teleportation network 
with arbitrary coherent signal states. 
Let us finally note that the generated tripartite entangled state, 
though being a Gaussian state with an always positive 
Wigner function, potentially 
exhibits nonlocal correlations between the three modes: 
it can be shown \cite{Loock01b} that this state 
violates three-party Bell-Mermin-Klyshko inequalities 
imposed by local realism \cite{Bell64a,Mermin90a,Klyshko93a}. 
However, instead of the relatively simple homodyne measurements, 
photon number parity measurements are required to detect 
this nonlocality \cite{Banaszek98a} which is unfeasible with current 
technology.

This work was supported by the MEXT and the MPHPT of Japan, and Research Foundation for Opto-Science and Technology.
PvL acknowledges financial support from the DFG under the Emmy-Noether programme.


\bibliography{aoki}

\end{document}